\documentclass{article}

\usepackage[nonatbib,preprint]{neurips_2024}

\usepackage[utf8]{inputenc} % allow utf-8 input
\usepackage[T1]{fontenc}    % use 8-bit T1 fonts
\usepackage{url}            % simple URL typesetting
\usepackage{booktabs}       % professional-quality tables
\usepackage{amsfonts}       % blackboard math symbols
\usepackage{nicefrac}       % compact symbols for 1/2, etc.
\usepackage{microtype}      % microtypography
\usepackage{xcolor}         % colors

\usepackage{bbding}
\usepackage{pifont}
\usepackage{amssymb}

\newcommand{\tabincell}[2]{\begin{tabular}{@{}#1@{}}#2\end{tabular}}
\usepackage{bm}
\usepackage{comment}
\usepackage{amsmath}
\usepackage{enumerate} 
\usepackage{enumitem}
\usepackage{mathrsfs}
\usepackage{makeidx}      
\usepackage{graphicx}       
\usepackage{multicol}       
\usepackage{subfig, wrapfig}
\usepackage{epsfig,amssymb,latexsym}
\usepackage{psfrag}
\usepackage[ruled,vlined]{algorithm2e}
\definecolor{lightgray}{rgb}{.91,.91,.91}
\definecolor{rouse}{rgb}{0.981,0.961,0.941}
\definecolor{deepred}{rgb}{0.698,0.133,0.133}
\definecolor{blue}{rgb}{0,0,1}

\usepackage{makecell}	
\usepackage{fancyhdr}
\usepackage{multirow}
\usepackage{rotating}
\usepackage{color, colortbl}

\usepackage[pagebackref=false,breaklinks=true,colorlinks,bookmarks=false,linkcolor=red,urlcolor=red,citecolor=blue]{hyperref}

\usepackage{amsthm}

\title{Dual Hyperspectral Mamba for Efficient Spectral Compressive Imaging}

\author{
 Jiahua~Dong\textsuperscript{1}\thanks{Equal Contributions.~~ $^\dag$Corresponding Author.}~~,~
 Hui Yin\textsuperscript{2$*$},~
 Hongliu Li\textsuperscript{3},~
 Wenbo Li\textsuperscript{4},~
 Yulun Zhang\textsuperscript{5$\dag$}, \\
 \textbf{Salman Khan\textsuperscript{1, 6},}~
 \textbf{Fahad Shahbaz Khan\textsuperscript{1, 7}} \\
 \textsuperscript{1}Mohamed bin Zayed University of Artificial Intelligence 
 ~~\textsuperscript{2}Hunan University \\ \textsuperscript{3}The Hong Kong Polytechnic University 
~~\textsuperscript{4}The Chinese University of Hong Kong \\
 \textsuperscript{5}Shanghai Jiao Tong University 
 ~~\textsuperscript{6}Australian National University
 ~~\textsuperscript{7}Link\"{o}ping University \\
}

\begin{document}

\maketitle

\begin{abstract}
Deep unfolding methods have made impressive progress in restoring 3D hyperspectral images (HSIs) from 2D measurements through convolution neural networks or Transformers in spectral compressive imaging. However, they cannot efficiently capture long-range dependencies using global receptive fields, which significantly limits their performance in HSI reconstruction. Moreover, these methods may suffer from local context neglect if we directly utilize Mamba to unfold a 2D feature map as a 1D sequence for modeling global long-range dependencies. To address these challenges, we propose a novel \underline{D}ual \underline{H}yperspectral \underline{M}amba (DHM) to explore both global long-range dependencies and local contexts for efficient HSI reconstruction. After learning informative parameters to estimate degradation patterns of the CASSI system, we use them to scale the linear projection and offer noise level for the denoiser (\emph{i.e.}, our proposed DHM). Specifically, our DHM consists of multiple dual hyperspectral S4 blocks (DHSBs) to restore original HSIs. Particularly, each DHSB contains a global hyperspectral S4 block (GHSB) to model long-range dependencies across the entire high-resolution HSIs using global receptive fields, and a local hyperspectral S4 block (LHSB) to address local context neglect by establishing structured state-space sequence (S4) models within local windows. Experiments verify the benefits of our DHM for HSI reconstruction. 
The source codes and models will be available at \url{https://github.com/JiahuaDong/DHM}.

\end{abstract}

\vspace{-2mm}
\section{Introduction}
\vspace{-2mm}

\begin{wrapfigure}{r}{0.37\textwidth}
\vspace{-10mm}
\begin{center} \hspace{-1.5mm}
\includegraphics[width=0.37\textwidth]{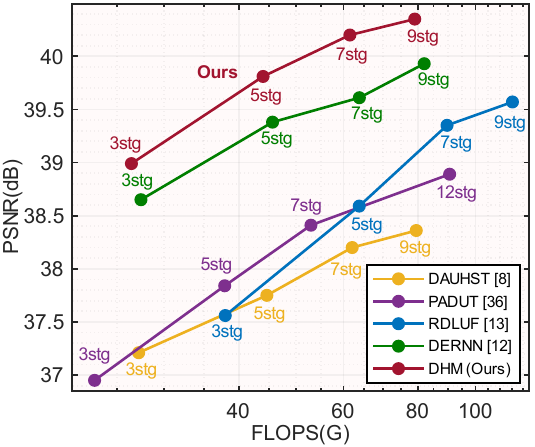}
\end{center}
\vspace{-4mm}
\caption{\small Comparisons of PSNR-FLOPS between our DHM and SOTA models.}
\vspace{-2mm}
\label{fig: teaser}
\end{wrapfigure}

Unlike standard RGB images with only three spectral bands, hyperspectral images (HSIs) \cite{5543780, 6297992, 7938635, 10.1126/science.228.4704.1147} comprise multiple contiguous bands, providing detailed spectral information for each pixel. In recent decades, HSIs have achieved remarkable successes in a wide range of applications such as remote sensing \cite{rs_1, 1323134, 7855724}, object detection \cite{10.1145/2185520.2185534, 1251148}, vehicle tracking \cite{7789671, 8014769, 7780774}, and medical image analysis \cite{mi_1, mi_2, mi_3}. With the development of compressive sensing theory, the coded aperture snapshot spectral imaging (CASSI) \cite{meng2020end, Gehm2007Express}, one of the snapshot compressive imaging systems \cite{7056558, Maxiao2021, QiaoMu23, Wagadarikar08}, has shown impressive performance in capturing HSIs at video rate. The CASSI system modulates HSI signals at various wavelengths, and mixes all modulated spectra to output a 2D compressed measurement. Then, numerous HSI reconstruction methods \cite{9351612, zhang2019computational, qiao2020deep} are developed to restore original HSIs from 2D compressed measurements (\emph{i.e.}, the CASSI inverse problem \cite{cai2022degradationaware}).

Different from natural image restoration, HSI reconstruction deals with substantially degraded measurements caused by uncertain system noise and spectral compression \cite{meng2020end, dernn_lnlt}. Thus, it is more challenging to learn underlying HSI properties than natural image restoration. Generally, existing HSI reconstruction methods can be mainly divided into four categories. To solve the CASSI inverse problem, model-based methods \cite{wang2016adaptive, zhang2019computational, 7328255} are heavily dependent on hand-crafted image priors (\emph{e.g.}, low-rank \cite{8481592} and sparsity \cite{Kittle2010Appl}), suffering from limited generalization capability. Some plug-and-play works \cite{zhang2021plug, ZhengSiming21Optica, 5543780} apply the pretrained denoiser into model-based methods \cite{qiao2020deep, yuan2020plug}, while end-to-end algorithms \cite{meng2020end, 9879297, lambda_Net} ignore the working mechanism of the CASSI system and instead model a brute-force projection from 2D compressed measurements to HSIs via convolutional neural networks (CNNs). Moreover, deep unfolding methods \cite{zhang2022herosnet, 9351612, 8954038, 9156942} introduce a multi-stage unfolding framework to iteratively learn a linear projection and a denoiser. They possess the interpretability of model-based methods \cite{9363502} as well as the powerful encoding capability of deep learning, thereby achieving state-of-the-art performance to lead the development of HSI reconstruction task.

Many deep unfolding methods \cite{zhang2022herosnet, 9363502} rely on CNNs as denoiser to capture local contexts, showing significant limitations in exploiting the crucial global contexts for HSI reconstruction. To tackle this issue, some works employ Transformers \cite{dosovitskiy2021an} to model wide-range dependencies \cite{dernn_lnlt, cai2022mask, cai2022coarse, cai2022degradation}, but the complexity is quadratic to the token size. Therefore, there is a trade-off between computation complexity and effective receptive fields, hindering these methods from exploring long-range dependencies, especially in high-resolution HSIs. Recently, structured state space sequence (S4) models \cite{gu2021efficiently, wang2023selective, mehta2022long} have emerged as a promising backbone to address the limitations of Transformers and CNNs. Then visual Mamba models \cite{zhu2024vision, wang2024mamba} introduce a cross-scan module to apply S4 models into vision tasks by unfolding 2D features as 1D array along four directions. It can use global receptive fields to capture long-range contexts while reducing the quadratic complexity to linear. However, existing Mamba models \cite{zhu2024vision, wang2024mamba} face a crucial challenge of local context neglect when directly applied to the high-resolution HSI reconstruction. Since Mamba unfolds a 2D feature map as a 1D sequence, spatially close pixels may end up being located at distant positions in the flattened sequences. The excessive distance among nearby pixels leads to the problem of local context neglect (\emph{i.e.}, significant loss of critical local textures), thereby degrading the performance of HSI reconstruction.

To resolve the above challenges, we develop a novel \underline{D}ual \underline{H}yperspectral \underline{M}amba (DHM) for efficient HSI reconstruction. Our DHM relies on structured state-space sequence (S4) models to reconstruct HSIs from 2D degraded measurements, which can capture both global long-range dependencies and local contexts with linear computation complexity. It is the first attempt to address HSI reconstruction via S4 models in the field of hyperspectral compressive imaging. After learning informative parameters from the physical mask and degraded measurement of the CASSI system, we feed them into multi-stage unfolding framework by scaling the linear projection and estimating noise level for the denoiser (\emph{i.e.}, our proposed DHM). The core component of our DHM is dual hyperspectral S4 block (DHSB), which is mainly composed of a \textit{global hyperspectral S4 block (GHSB)} and a \textit{local hyperspectral S4 block (LHSB)}. More specifically, the GHSB focuses on understanding global long-range dependencies by modeling discrete state-space equation on the entire high-resolution HSIs, which can effectively balance computation complexity and global receptive fields. Besides, the LHSB aims to surmount the challenge of local context neglect by constructing S4 models within different local windows. As shown in Fig.~\ref{fig: teaser}, experiments shows that our DHM significantly surpasses existing HSI reconstruction methods. The novel contributions of our paper are listed as follows:

$\bullet$ We propose a new Dual Hyperspectral Mamba (DHM) for HSI reconstruction, capable of capturing both global long-range dependencies and local contexts with linear computational complexity. To our best knowledge, our DHM is the first Mamba-based deep unfolding method for HSI reconstruction. 

$\bullet$ We develop a global hyperspectral S4 block (GHSB) to explore long-range dependencies across the entire high-resolution HSIs using global receptive fields, while design a local hyperspectral S4 block (LHSB) to tackle local context neglect by constructing S4 models within different local windows. 

$\bullet$ We conduct comprehensive experiments to illustrate that our DHM significantly surpasses SOTA deep unfolding methods, while requiring lower model size and computational complexity. 

\vspace{-4mm}
\section{Related Work}
\vspace{-2mm}
\textbf{Hyperspectral Image Reconstruction:}
Traditional model-based HSI reconstruction methods \cite{7780774, 7328255, wang2016adaptive, 7532817, zhang2019computational} utilize hand-crafted priors such as sparsity \cite{Kittle2010Appl}, total total variation \cite{7532817} and low-rank constraint to address the CASSI inverse problem. Unfortunately, they highly rely on manual parameter tuning, leading to unsatisfactory reconstruction performance. In light of this, some plug-and-play methods \cite{zhang2021plug, 5543780, lai2022deep} focus on integrating convex optimization with the pretrained denoising networks for HSI reconstruction. They have limited generalization performance due to the overreliance on pretrained denoiser.
Besides, end-to-end (E2E) algorithms \cite{cai2022mask, 9879297, meng2020end} rely on convolutional neural networks (CNNs) \cite{What_Transferred_Dong_CVPR2020, 9616392_dong} or Transformers \cite{dosovitskiy2021an} to learn a brute-force projection function for HSI restoration. They can improve the HSI reconstruction performance but lack robustness and interpretability. To address these limitations, deep unfolding methods \cite{cai2022degradationaware, 9351612, huang2021deep, 9009497, 9156942} are developed to restore HSI cubes from 2D compressed measurements via a multi-stage framework, showcasing the interpretability and strong encoding ability. \cite{9009497, zhang2022herosnet, 9156942} employ CNNs to estimate degradation patterns, showing limitations to explore long-range contexts. After Cai \emph{et al.} \cite{cai2022degradationaware} employ Transformer to capture non-local dependencies, many Transformer-based methods \cite{huang2023deep, dong2023residual, li2023pixel, dernn_lnlt, wu2023latent} are proposed to design the denoisers. 
However, the above methods suffer from a trade-off between computation complexity and effective receptive fields, preventing them from understanding long-range dependencies with global receptive fields to achieve better HSI reconstruction performance.

\textbf{State Space Models} (SSMs) \cite{gu2021efficiently, gu2021combining, smith2022simplified, koller2009probabilistic} have attracted increasing attention recently due to their capability to linearly scale with sequence length in the long-range dependency modeling. After structured state space sequence (S4) model \cite{gu2021efficiently} shows impressive performance on long-range sequence modeling tasks, S5 model \cite{smith2022simplified} introduces an efficient parallel scan and a general MIMO SSM based on S4. Then \cite{fu2022hungry, mehta2022long} are proposed to alleviate the performance gap between Transformers and SSMs.
Mamba \cite{gu2023mamba}, an enhanced SSM with efficient hardware design and a selective mechanism, has surpassed Transformer in natural language processing \cite{islam2023efficient, nguyen2022s4nd}. Due to its ability in modeling long-range dependencies with linear complexity, Mamba has been widely applied to diverse vision tasks, such as image/video understanding \cite{liu2024vmamba, yue2024medmamba, li2024videomamba} and biomedical image analysis \cite{ma2024u}. However, these Mamba models \cite{gu2023mamba, yue2024medmamba, li2024videomamba, liu2024vmamba, nguyen2022s4nd} 'may face the challenge of local context neglect (\emph{i.e.}, substantial loss of critical local textures), when directly applied to the high-resolution HSI reconstruction task.

\vspace{-2mm}
\section{The Proposed Model}
\vspace{-2mm}
\subsection{The CASSI System}
\vspace{-2mm}
\begin{wrapfigure}{r}{0.65\textwidth}
\vspace{-15mm}
\begin{center} \hspace{-1.5mm}
\includegraphics[width=0.65\textwidth]{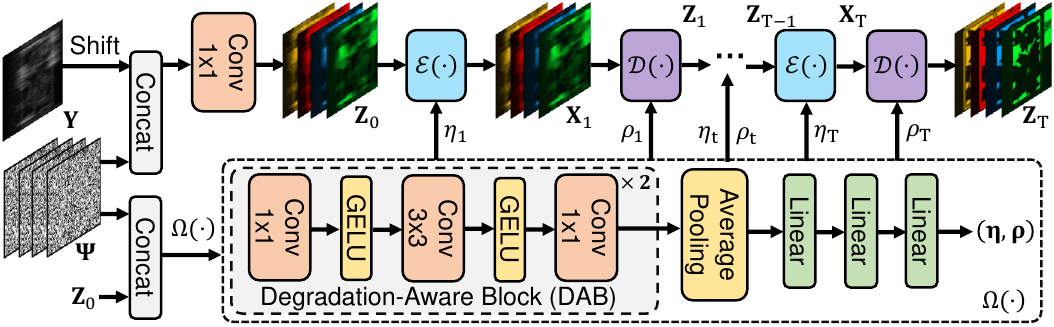}
\end{center}
\vspace{-4mm}
\caption{\small Our unfolding framework with $T$ iterative stages.}
\vspace{-2mm}
\label{fig: DU_pipeline}
\end{wrapfigure}

\textbf{Degradation Model:}
In the coded aperture snapshot spectral imaging (CASSI) system \cite{4385788, meng2020end, Gehm2007Express}, the camera can capture the vectorized degraded measurement $\mathbf{Y}\in \mathbb{R}^{\xi}$, where $\xi = H(\delta_s(N_\omega-1) + W)$. $N_\omega, \delta_s, H$ and $W$ represent the number of wavelengths, shifting step of dispersion, height and width in hyperspectral images (HSIs), respectively. As introduced in \cite{cai2022degradationaware}, after vectorizing the shifted HSI signal as $\mathbf{X} \in\mathbb{R}^{\xi N_\omega}$, we express the degradation model of the CASSI system as follows:
\begin{align}
\mathbf{Y} = \boldsymbol{\Psi} \mathbf{X} + \boldsymbol{\epsilon},
\label{eq: CASSI_system}
\end{align}
where $\boldsymbol{\epsilon} \in \mathbb{R}^{\xi}$ is the vectorized imaging noise on $\mathbf{Y}$. $\mathbf{\Psi} \in\mathbb{R}^{\xi\times\xi N_\omega}$ indicates the sparse and fat sensing matrix which is determined via the physical mask in the CASSI system \cite{1614066, 8830474}. 
Given $\mathbf{\Psi}$ and $\mathbf{Y}$ in the CASSI system, the goal of HSI
reconstruction is to restore HSI signal $\mathbf{X}$ by removing the imaging noise $\boldsymbol{\epsilon}$.

\textbf{Estimation of Degradation Patterns:} 
As analyzed in previous deep unfolding methods \cite{cai2022degradationaware, dong2023residual, dernn_lnlt, wu2023latent}, the estimation of degradation patterns is crucial to improve HSI reconstruction performance in the multi-stage unfolding framework, by adaptively scaling linear projection and offering information about imaging noise $\boldsymbol{\epsilon}$ for the denoiser. Thus, motivated by \cite{cai2022degradationaware, dong2023residual}, we use maximum a posteriori (MAP) theory to restore original HSI signal $\mathbf{X}$ in Eq.~\eqref{eq: CASSI_system} via optimizing the following energy function:
\begin{align}
\widehat{\mathbf{X}} = \arg \min_\mathbf{X} \frac{1}{2} \|\mathbf{Y-\mathbf{\Psi X}}\|^2 + \lambda\mathcal{R}(\mathbf{X}),
\label{eq: objective_map_x}
\end{align}
where $\mathcal{R}(\mathbf{X})$ denotes the prior term about $\mathbf{X}$, and $\lambda$ is the hyperparameter to balance the importance of prior term. In order to solve Eq.~\eqref{eq: objective_map_x}, we define an auxiliary variable as $\mathbf{Z} =\mathbf{X}\in\mathbb{R}^{\xi N_\omega}$, and then utilize the half-quadratic splitting algorithm to minimize the following loss $\mathcal{L}_{\mathrm{HSI}}$:
\begin{align}
\mathcal{L}_{\mathrm{HSI}}= \frac{1}{2} \|\mathbf{Y-\mathbf{\Psi X}}\|^2 + \lambda\mathcal{R}(\mathbf{Z}) + \frac{\eta}{2} \| \mathbf{Z} - \mathbf{X}\|^2,
\label{eq: optimization_loss}
\end{align}
where $\eta$ is a penalty parameter. We decouple $\mathbf{X}$ and $\mathbf{Z}$ into two iterative subproblems to solve Eq.~\eqref{eq: optimization_loss}:
\begin{align}
\mathbf{X}_t = \arg\min_{\mathbf{X}} \|\mathbf{Y} - \mathbf{\Psi X}\|^2 + \eta \|\mathbf{X} - \mathbf{Z}_{t-1}\|^2, ~~\mathbf{Z}_t = \arg\min_{\mathbf{Z}} + \frac{\eta}{2} \|\mathbf{Z} - \mathbf{X}_{t} \|^2 + \lambda\mathcal{R}(\mathbf{Z}),
\label{eq: subproblem}
\end{align}
where $t=1, \cdots, T$ denotes the iterative stage index in the multi-stage unfolding framework, as shown in Fig.~\ref{fig: DU_pipeline}. Since the subproblem of solving $\mathbf{X}$ in Eq.~\eqref{eq: subproblem} is a quadratic regularized least-squares problem, we can derive its closed solution as 
$\mathbf{X}_t = (\boldsymbol{\Psi}^\top \boldsymbol{\Psi} + \eta\mathbf{I})^{-1} (\boldsymbol{\Psi}^\top \mathbf{Y} + \eta \mathbf{Z}_{t-1})$.
Considering the high computational overhead of $(\boldsymbol{\Psi}^\top \boldsymbol{\Psi} + \eta\mathbf{I})^{-1}$ brought by the fat sensing matrix $\boldsymbol{\Psi}\in\mathbb{R}^{\xi\times\xi N_\omega}$, we resort to the matrix inversion formula to simplify it: $(\boldsymbol{\Psi}^\top \boldsymbol{\Psi} + \eta\mathbf{I})^{-1} = {\eta}^{-1}\mathbf{I} - {\eta}^{-1}\boldsymbol{\Psi}^\top (\boldsymbol{\Psi} {\eta}^{-1} \boldsymbol{\Psi}^\top + \mathbf{I})^{-1} \boldsymbol{\Psi} \eta^{-1}$. As a result, we can reformulate the closed solution of $\mathbf{X}$ in Eq.~\eqref{eq: subproblem} as follows:
\begin{align}
\!\!\!\mathbf{X}_t \!=\! \mathbf{Z}_{t-1} + \eta^{-1}\boldsymbol{\Psi}^\top \mathbf{Y} \!-\!
\eta^{-1} \boldsymbol{\Psi}^\top (\boldsymbol{\Psi} {\eta}^{-1} \boldsymbol{\Psi}^\top \!\!+\! \mathbf{I})^{-1} \boldsymbol{\Psi} \mathbf{Z}_{t-1} \!-\! 
\eta^{-2} \boldsymbol{\Psi}^\top (\boldsymbol{\Psi} {\eta}^{-1} \boldsymbol{\Psi}^\top \!\!+\! \mathbf{I})^{-1} \boldsymbol{\Psi} \boldsymbol{\Psi}^\top\mathbf{Y}.\!\!\!
\label{eq: solution_X_final}
\end{align}
As introduced in \cite{cai2022coarse, cai2022degradationaware}, $\boldsymbol{\Psi} \boldsymbol{\Psi}^\top = \boldsymbol{\mathrm{diag}}\{\psi_1, \psi_2, \cdots, \psi_\xi\}$ is a diagonal matrix in the CASSI system. After defining $\boldsymbol{\psi} = [\psi_1, \psi_2, \cdots, \psi_\xi] \in\mathbb{R}^{\xi}$, we plug $\boldsymbol{\Psi} \boldsymbol{\Psi}^\top = \boldsymbol{\mathrm{diag}}\{\psi_1, \psi_2, \cdots, \psi_\xi\}$ into Eq.~\eqref{eq: solution_X_final}: 
\begin{align}
\mathbf{X}_t = \mathbf{Z}_{t-1} + \boldsymbol{\Psi}^\top( (\mathbf{Y} - \boldsymbol{\Psi}\mathbf{Z}_{t-1}) \otimes (\eta+\boldsymbol{\psi})^{-1})^\top, 
\label{eq: solution_X_final_simplified}
\end{align}
where $\otimes$ is the element-wise multiplication. Since $\boldsymbol{\psi}$ is precomputed and stored in $\boldsymbol{\Psi} \boldsymbol{\Psi}^\top$, the value of $\eta$ in Eq.~\eqref{eq: solution_X_final_simplified} can affect the output of each iterative stage in the multi-stage unfolding framework. To eliminate negative influence of manually determining $\eta$, we set $\eta$ to be learnable in the multi-stage framework, and denote $\eta_t$ as the value of $\eta$ at the $t$-th iterative stage. Besides, we also define a learnable parameter $\lambda_t$ at the $t$-th stage, and express the subproblem of solving $\mathbf{Z}_t$ in Eq.~\eqref{eq: subproblem} as: 
\begin{align}
\mathbf{Z}_t = \arg\min_{\mathbf{Z}} \frac{1}{2(\sqrt{\lambda_t/\eta_t})^2} \|\mathbf{Z} - \mathbf{X}_{t} \|^2 + \mathcal{R}(\mathbf{Z}). 
\label{eq: subproblem_Z_final}
\end{align}
In Eq.~\eqref{eq: subproblem_Z_final}, the subproblem of solving $\mathbf{Z}_t$ is equivalent to denoising the image $\mathbf{X}_t$ with a Gaussian noise level of $\sqrt{\lambda_t/\eta_t}$, according to Bayesian probability \cite{7744574}. Given $\boldsymbol{\eta} = [\eta_1, \cdots, \eta_T] \in\mathbb{R}^T$ and $\boldsymbol{\rho} = [\eta_1/\lambda_1, \cdots, \eta_T/\lambda_T] \in\mathbb{R}^T$, we can introduce the following iterative optimization scheme to estimate degradation patterns of the CASSI system and reconstruct original HSI signal $\mathbf{X}$ in Eq.~\eqref{eq: CASSI_system}:
\begin{align}
(\boldsymbol{\eta}, \boldsymbol{\rho}) = \Omega(\boldsymbol{\Psi}, \mathbf{Y}), ~
\mathbf{X}_t = \mathcal{E}(\mathbf{Z}_{t-1}, \boldsymbol{\Psi}, \mathbf{Y}, \eta_t),~
\mathbf{Z}_t = \mathcal{D}(\mathbf{X}_t, \rho_t),
\label{eq: iterative_updating_scheme}
\end{align}
where $\Omega(\cdot)$ is the parameter learner. $\mathcal{E}(\cdot)$ is equivalent to Eq.~\eqref{eq: solution_X_final_simplified}, which is a linear projection used for mapping $\mathbf{Z}_{t-1}$ to $\mathbf{X}_{t}$. $\mathcal{D}(\cdot)$ indicates the Gaussian denoiser to solve Eq.~\eqref{eq: subproblem_Z_final}. As shown in Fig.~\ref{fig: DU_pipeline}, we depict our unfolding framework with $T$ iterative training stages to restore original HSI signal $\mathbf{X}$ in Eq.~\eqref{eq: CASSI_system}. Specifically, we first concatenate the given sensing matrix $\boldsymbol{\Psi}$ and compressed measurement $\mathbf{Y}$, and input it into a convolution block to initialize $\mathbf{Z}_0$. At the $t$-th ($t=1, \cdots, T$) stage, the parameter learner $\Omega(\cdot)$ contains two degradation-aware blocks (DABs), an average pooling layer and three fully connected layers to encode $\mathbf{Z}_0$ and $\boldsymbol{\Psi}$, and then outputs learnable parameters $(\boldsymbol{\eta}, \boldsymbol{\rho})$. The DAB has three convolution layers and two GELU functions.
Then $\mathcal{E}(\cdot)$ and $\mathcal{D}(\cdot)$ use the parameters $(\boldsymbol{\eta}, \boldsymbol{\rho})$ to iteratively update $\mathbf{X}_t$ and $\mathbf{Z}_t$ in Eq.~\eqref{eq: iterative_updating_scheme} until the $T$-th stage. Particularly, $(\boldsymbol{\eta}, \boldsymbol{\rho})$ learned by $\Omega(\cdot)$ can effectively scale the
linear projection in Eq.~\eqref{eq: solution_X_final_simplified}, while offering accurate noise level for the denoiser $\mathcal{D}(\cdot)$ to solve Eq.~\eqref{eq: subproblem_Z_final}. In the CASSI system, they are essential to estimate the ill-posedness degree and degradation patterns, thereby substantially improving HSI reconstruction performance.

\vspace{-5pt}
\subsection{Dual Hyperspectral Mamba (DHM)}
\vspace{-5pt}
Generally, existing deep unfolding methods \cite{cai2022degradation, dernn_lnlt, wu2023latent, dong2023residual} mainly utilize CNNs or Transformers to design the denoiser $\mathcal{D}(\cdot)$. However, these methods struggle to capture long-range dependencies using global receptive fields, thereby limiting their HSI reconstruction performance. Besides, directly applying Mamba to high-resolution HSI reconstruction suffers from local context neglect (\emph{i.e.}, substantial loss of critical local details). To resolve the above challenges, we develop a novel Dual Hyperspectral Mamba (DHM) as the denoiser $\mathcal{D}(\cdot)$ in Eq.~\eqref{eq: iterative_updating_scheme}. Our DHM uses global receptive fields to model long-range dependencies while tackling local context neglect via capturing local contexts.

\begin{figure*}[t]
\centering
\includegraphics[width=.99999\linewidth]
{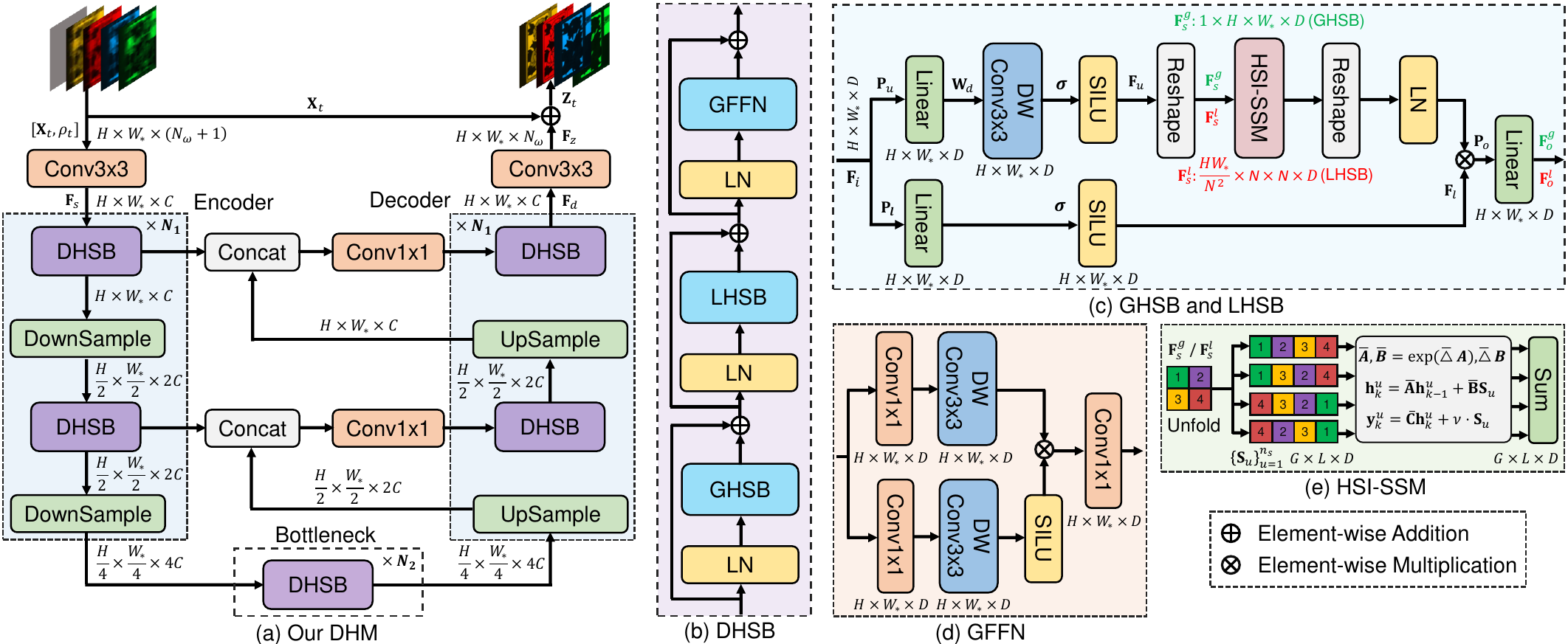}
\vspace{-6mm}
\caption{Algorithmic pipeline of our DHM. (a) Architecture of our DHM at the $t$-th iterative stage. (b) Each DHSB is composed of a GHSB, a LHSB, a GFFN and three LN layers. (c) Diagram of the GHSB and LHSB modules. (d) Components of the GFFN. (e) Design of the HSI-SSM. }
\label{fig: model_overview}
\vspace{-2mm}
\end{figure*}

Fig.~\ref{fig: model_overview}\textcolor{red}{a} shows the architecture of our DHM (\emph{i.e.}, the denoiser $\mathcal{D}(\cdot)$) at the $t$-th ($t=1, \cdots, T$) iterative stage in Fig.~\ref{fig: DU_pipeline}. Specifically, given the scalar $\rho_t$ and $\mathbf{X}_t \in\mathbb{R}^{H\times 
W_*\times N_\omega}$ at the $t$-th stage, we first reshape $\rho_t$ to $\mathbb{R}^{H\times W_*}$, and concatenate $\mathbf{X}_t$ with the reshaped $\rho_t$ to extract shallow feature $\mathbf{F}_s\in\mathbb{R}^{H\times 
W_*\times C}$ via a convolutional layer, where $W_*=\delta_s(N_\omega-1)+W$, and $C$ is the feature dimension. Then we forward $\mathbf{F}_s$ to the encoder, bottleneck and decoder to obtain the deep feature $\mathbf{F}_d \in\mathbb{R}^{H\times W_*\times C}$. The encoder and decoder comprise $N_1$ pairs of dual hyperspectral S4 block (DHSB) and the resizing module, while the bottleneck only has $N_2$ DHSBs. In Fig.~\ref{fig: model_overview}\textcolor{red}{a}, we visualize the pipeline of our DHM when $N_1=2$ and $N_2=1$ for better demonstration. In Fig.~\ref{fig: model_overview}\textcolor{red}{b}, the DHSB includes a global hyperspectral S4 block (GHSB), a local hyperspectral S4 block (LHSB), a gated feed-forward network (GFFN) and three layer normalization (LN). Fig.~\ref{fig: model_overview}\textcolor{red}{c} presents the components of GHSB and LHSB, which are the two most important modules in our DHM. Apart from the reshaping operation, they have the same architectures. Particularly, the GHSB can use global receptive fields to model long-range dependencies, and the LHSB aims to address local context neglect by constructing structured state space sequence (S4) model within local windows. Besides, Fig.~\ref{fig: model_overview}\textcolor{red}{d} shows the design of GFFN module. Then we perform a convolution operation on $\mathbf{F}_d$ to obtain $\mathbf{F}_z \in\mathbb{R}^{H\times W_*\times N_\omega}$. Finally, we sum $\mathbf{X}_t$ and $\mathbf{F}_z$ to generate the denoised image $\mathbf{Z}_t \in\mathbb{R}^{H\times W_*\times N_\omega}$ at the $t$-th iterative stage. In the following subsections, we introduce the detailed components of the GHSB and LHSB.

\textbf{Global Hyperspectral S4 Block (GHSB)} constructs S4 model on the entire high-resolution HSIs to capture global contexts using global receptive fields. As shown in Fig.~\ref{fig: model_overview}\textcolor{red}{c}, we forward a given feature $\mathbf{F}_i \in\mathbb{R}^{H\times W_*\times D}$ into two branches, where $D=\{C, 2C, 4C\}$ denotes the feature dimensions at different levels of encoder, bottleneck and decoder. Specifically, the upper branch encodes $\mathbf{F}_{i}$ to $\mathbf{F}_u = \boldsymbol{\sigma}(\mathbf{P}_u (\mathbf{W}_d(\mathbf{F}_i))) \in\mathbb{R}^{H\times W_*\times D}$ via a linear projection $\mathbf{P}_u(\cdot)$, a depth-wise convolution $\mathbf{W}_d(\cdot)$ and a SILU activation function $\boldsymbol{\sigma}(\cdot)$. Then we reshape $\mathbf{F}_u$ as $\mathbf{F}_s^g \in\mathbb{R}^{1\times H\times W_*\times D}$, and input it into the $\mathrm{HSI}\text{-}\mathrm{SSM}(\cdot)$ to model long-range dependencies using global receptive fields. As a result, we can formulate the output feature $\mathbf{F}_o^g \in\mathbb{R}^{H\times W_*\times D}$ of the GHSB module as follows:
\begin{align}
\mathbf{F}_o^g = \mathbf{P}_o\big(\mathrm{LN}(\mathrm{RS}(\mathrm{HSI}\text{-}\mathrm{SSM}(\mathbf{F}_s^g))) \otimes \mathbf{F}_l \big),
\label{eq: local_mamba}
\end{align}
where $\otimes$ denotes the element-wise multiplication. $\mathbf{F}_l= \boldsymbol{\sigma}(\mathbf{P}_l (\mathbf{F}_i)) \in \mathbb{R}^{H\times W_*\times D}$ denotes the output of lower branch in Fig.~\ref{fig: model_overview}\textcolor{red}{c}, and $\mathbf{P}_l(\cdot)$ is the linear mapping. $\mathrm{LN}(\cdot)$ is the layer normalization (LN), $\mathrm{RS}(\cdot)$ can reshape the given feature to $\mathbb{R}^{H\times W_*\times D}$, and $\mathbf{P}_o$ is the linear projection to obtain $\mathbf{F}_o^g$. Moreover, $\mathrm{HSI}\text{-}\mathrm{SSM}(\cdot)$ denotes the proposed hyperspectral image state space module (HSI-SSM).

\textbf{HyperSpectral Image State Space Module (HSI-SSM)}
can model long-range cross-pixel interactions to explore global contexts of $\mathbf{F}_i$ using global receptive fields. As shown in Fig.~\ref{fig: model_overview}\textcolor{red}{e}, given the input feature $\mathbf{F}_s^g \in\mathbb{R}^{1\times H\times W_*\times D}$, we unfold the entire hyperspectral image (HSI) that includes $H\times W_*$ pixels, into four one-dimensional sequences with a size of $HW_*$, by scanning these pixels along four distinct traversal paths: from the top-left to the bottom-right, from the top-right to the bottom-left, from the bottom-right to the top-left, and from the bottom-left to the top-right. We denote four sequence features as $\{\mathbf{S}_u \in\mathbb{R}^{G\times L\times D} \}_{u=1}^{n_s}$, where $n_s=4, G=1$, and $L=HW_*$ denotes the sequence length in the GHSB. Motivated by Mamba \cite{gu2023mamba, liu2024vmamba, wang2024mamba}, we construct some enhanced discrete state space equations on the $u$-th ($u=1, \cdots, n_s$) sequence feature $\mathbf{S}_u$. Specifically, after defining the learnable variables: $\mathbf{A} \in\mathbb{R}^{D\times D_s}$ and $\mathbf{E} \in\mathbb{R}^{G\times L\times D}$, we can formulate some continuous parameters such as $\mathbf{B} \in\mathbb{R}^{G\times L\times D_s}, \mathbf{C} \in\mathbb{R}^{G\times L\times D_s}$ and a timescale parameter $\triangle \in\mathbb{R}^{G\times L\times D}$ as:
\begin{align}
\mathbf{B} = \mathbf{P}_b(\mathbf{S}_u), ~~\mathbf{C} = \mathbf{P}_c(\mathbf{S}_u), ~~\triangle = \tau_\triangle(\mathbf{E}+ \mathbf{P}_\triangle(\mathbf{S}_u)),
\label{eq: continuous_parameters}
\end{align}
where $D_s$ is the latent feature dimension, and $\tau_\triangle(\cdot)$ is the softplus activation function. $\mathbf{P}_b(\cdot), \mathbf{P}_c(\cdot)$ and $\mathbf{P}_\triangle(\cdot)$ are the linear projection matrices. Inspired by the zero-order hold (ZOH) discretization rule \cite{gu2023mamba}, we reshape the parameter $\triangle$ as $\overline{\triangle}\in\mathbb{R}^{G\times L\times D\times 1}$, and utilize it to
transform the continuous parameters $\mathbf{A}$ and $\mathbf{B}$ into the discrete parameters $\overline{\mathbf{A}} \in\mathbb{R}^{G\times L\times D\times D_s}$ and $\overline{\mathbf{B}} \in\mathbb{R}^{G\times L\times D\times D_s}$:
\begin{align}
\overline{\mathbf{A}} = \exp(\overline{\triangle}\mathbf{A}), ~\overline{\mathbf{B}}= (\overline{\triangle} \mathbf{A})^{-1}(\exp(\overline{\triangle}\mathbf{A}) - \mathbf{I}) \cdot \overline{\triangle}\mathbf{B}. 
\label{eq: discrete_A_B}
\end{align}
After obtaining the discrete $\overline{\mathbf{A}}$ and $\overline{\mathbf{B}}$ via Eq.~\eqref{eq: discrete_A_B}, we reshape the parameter $\mathbf{C}$ as $\overline{\mathbf{C}}\in\mathbb{R}^{G\times L\times D_s\times 1}$, and formulate the semantic encoding of $\mathbf{S}_u$ as the form of recurrent neural networks (RNNs) to extract a new sequence feature $\mathbf{y}_k^u \in\mathbb{R}^{G\times L\times D}$. Then we denote $\mathbf{h}_{k-1}^u, \mathbf{h}_k^u \in\mathbb{R}^{G\times L\times D\times D_s}$ as the latent features of the $(k\!-\!1)$-th and $k$-th hidden states in the RNNs, and define $\mathbf{y}_k^u$ as follows:
\begin{align}
\mathbf{h}_k^u = \overline{\mathbf{A}} \mathbf{h}_{k-1}^u + \overline{\mathbf{B}} \mathbf{S}_u, ~~\mathbf{y}_k^u = \overline{\mathbf{C}}\mathbf{h}_k^u + \nu \cdot\mathbf{S}_u,
\label{eq: mamba_HSI}
\end{align}
where $\nu$ denotes the scale parameter. Inspired by \cite{gu2023mamba}, we use the broadcasting mechanism to match the dimensions of different matrices for matrix multiplication operations in Eqs.~\eqref{eq: discrete_A_B}\eqref{eq: mamba_HSI}. Then we merge all sequence features $\{\mathbf{y}_k^u\}_{u=1}^{n_s}$ to get the final output map $\mathbf{y} = \sum_{u=1}^{n_s} \mathbf{y}_k^u$ of the HSI-SSM. In the GHSB, we utilize the HSI-SSM to encode the entire high-resolution HSI in a recursive manner. It can explore long-range dependencies of the input feature $\mathbf{F}_i$ using global receptive fields.

\textbf{Local Hyperspectral S4 Block (LHSB)} aims to explore local contexts within position-specific windows. Different from the GHSB that uses the HSI-SSM to unfold and scan the entire high-resolution HSI containing $H\times W_*$ pixels, the LHSB scans each local window, including $N\times N$ pixels, to capture local contexts. Specifically,  as shown in Fig.~\ref{fig: model_overview}\textcolor{red}{c}, after encoding the given feature $\mathbf{F}_i$ to $\mathbf{F}_u \in\mathbb{R}^{H\times W_*\times D}$ via the upper branch, we partition $\mathbf{F}_u$ to $\nicefrac{H}{N} \times \nicefrac{W_*}{N}$ non-overlapping windows, then reshape it as $\mathbf{F}_s^l \in\mathbb{R}^{\nicefrac{HW_*}{N^2}\times N\times N\times D}$, and input $\mathbf{F}_s^l$ into the HSI-SSM, where $\nicefrac{HW_*} {N^2}$ denotes the number of windows and each window includes $N^2$ pixels. In the HSI-SSM, we flatten each window including $N^2$ pixels and scan them along four distinctive directions to obtain four sequence features $\{\mathbf{S}_u \in\mathbb{R}^{G\times L\times D} \}_{u=1}^{n_s}$. Note that we set $G=\nicefrac{HW_*}{N^2}$ and $L=N^2$ in the LHSB, which are different from the GHSB. After encoding each sequence $\{\mathbf{S}_u\}_{u=1}^{n_s}$ under a recursive manner to get $\{\mathbf{y}_k^u\}_{u=1}^{n_s}$, we sum them to get the output map $\mathbf{y} \in\mathbb{R}^{G\times L\times D}$ of the HSI-SSM. The LHSB can capture local contexts of HSI by encoding different local windows of the  given feature $\mathbf{F}_i$ in a recursive manner. Thus, we formulate the final feature $\mathbf{F}_o^l \in\mathbb{R}^{H\times W_*\times D}$ outputted by the LHSB as follows:
\begin{align}
\mathbf{F}_o^l = \mathbf{P}_o\big(\mathrm{LN}(\mathrm{RS}(\mathrm{HSI}\text{-}\mathrm{SSM}(\mathbf{F}_s^l))) \otimes \mathbf{F}_l \big).
\label{eq: global_mamba}
\end{align}
\textbf{Optimization:}
As shown in Fig.~\ref{fig: DU_pipeline}, we utilize $\mathcal{E}(\cdot)$ and $\mathcal{D}(\cdot)$ (\emph{i.e.}, our DHM) to iteratively update $\mathbf{X}_t$ and $\mathbf{Z}_t$ in Eq.~\eqref{eq: iterative_updating_scheme} until the $T$-th stage. After getting $\mathbf{Z}_T$ at the $T$-th stage, we follow \cite{dernn_lnlt, dong2023residual} to train our DHM by minimizing the Charbonnier loss between the groundtruth and reconstructed HSI $\mathbf{Z}_T$.

\vspace{-3mm}
\section{Experiments} \label{sec: experiments}
\vspace{-2mm}
\subsection{Implementation Details}
\vspace{-5pt}
For fair comparisons, we set exactly the same experimental configurations with existing HSI reconstruction methods \cite{cai2022degradation, wu2023latent, 9741335, dernn_lnlt, 9857277, 9879297} to validate the effectiveness of our DHM. Following the settings of \cite{9879297, meng2020gap, 9857277, meng2020end}, we perform spectral interpolation on the original HSIs and choose a wide spectral range from 450 nm to 650 nm for comparisons on both the simulation and real datasets. The simulation dataset is composed of two subsets: KAIST \cite{kaist} and CAVE \cite{cave}. We employ the CAVE subset to train our DHM, and select 10 HSIs from the KAIST to evaluate performance. Moreover, the real dataset \cite{meng2020end} consists of five HSI cubes, which are captured by the practical CASSI system \cite{meng2020end}.

During training, we employ the Adam optimizer \cite{adam} to train all variants of our DHM on a single NVIDIA A100 GPU, where initial learning rate is $1.0\times 10^{-3}$, and the training epoches are set to 300. Following \cite{cai2022degradation, wu2023latent, dernn_lnlt, 9879297}, we randomly crop HSI cubes to $256\times 256 \times 28$ for simulation dataset, and $660\times 660 \times 28$ for real dataset. The shifting step of dispersion in the CASSI system is set to $\delta_s=2$. Moreover, we set $C=28, N=8, N_1=2, N_2=1$ and $D=D_s$ in this paper. Motivated by baseline HSI reconstruction methods \cite{dernn_lnlt, dong2023residual}, we share the network weights of our DHM across different stages, and use exactly the same data augmentation to train our DHM.

\begin{table*}[t]
\caption{Performance of our DHM and other comparison methods on the simulation dataset with 10 scenes (S1$\sim$S10). In each cell, the upper and lower entries report PSNR and SSIM, respectively.}
\vspace{-2mm}
\centering
\resizebox{0.99999\textwidth}{!}
{
    % \centering
    % \begin{tabular}{c|c|c|c|c|c|c|>{\columncolor{lightgray}}c}
    \begin{tabular}{l|cc|cccccccccc|c}
        \toprule
        % \rowcolor{lightgray}
        \makecell[c]{Comparison Methods}
        &\#Params
        &GFLOPS
        & ~~~S1~~~
        & ~~~S2~~~
        & ~~~S3~~~
        & ~~~S4~~~
        & ~~~S5~~~
        & ~~~S6~~~
        & ~~~S7~~~
        & ~~~S8~~~
        & ~~~S9~~~
        & ~~~S10~~~
        & ~~~Avg~~~
        \\
        \midrule
        TwIST \cite{4358846}
        & - 
        & -
        &\tabincell{c}{25.16\\0.700}
        &\tabincell{c}{23.02\\0.604}
        &\tabincell{c}{21.40\\0.711}
        &\tabincell{c}{30.19\\0.851}
        &\tabincell{c}{21.41\\0.635}
        &\tabincell{c}{20.95\\0.644}
        &\tabincell{c}{22.20\\0.643}
        &\tabincell{c}{21.82\\0.650}
        &\tabincell{c}{22.42\\0.690}
        &\tabincell{c}{22.67\\0.569}
        &\tabincell{c}{23.12\\0.669}
        \\
        % \midrule
        % GAP-TV \cite{7532817}
        % & - 
        % & -
        % &\tabincell{c}{26.82\\0.754}
        % &\tabincell{c}{22.89\\0.610}
        % &\tabincell{c}{26.31\\0.802}
        % &\tabincell{c}{30.65\\0.852}
        % &\tabincell{c}{23.64\\0.703}
        % &\tabincell{c}{21.85\\0.663}
        % &\tabincell{c}{23.76\\0.688}
        % &\tabincell{c}{21.98\\0.655}
        % &\tabincell{c}{22.63\\0.682}
        % &\tabincell{c}{23.10\\0.584}
        % &\tabincell{c}{24.36\\0.669}
        % \\
        % \midrule
        % DeSCI \cite{8481592}
        % & - 
        % & -
        % &\tabincell{c}{27.13\\0.748}
        % &\tabincell{c}{23.04\\0.620}
        % &\tabincell{c}{26.62\\0.818}
        % &\tabincell{c}{34.96\\0.897}
        % &\tabincell{c}{23.94\\0.706}
        % &\tabincell{c}{22.38\\0.683}
        % &\tabincell{c}{24.45\\0.743}
        % &\tabincell{c}{22.03\\0.673}
        % &\tabincell{c}{24.56\\0.732}
        % &\tabincell{c}{23.59\\0.587}
        % &\tabincell{c}{25.27\\0.721}
        % \\
        \midrule
        $\lambda$-Net \cite{lambda_Net}
        & 62.64M
        & 117.98
        &\tabincell{c}{30.10\\0.849}
        &\tabincell{c}{28.49\\0.805}
        &\tabincell{c}{27.73\\0.870}
        &\tabincell{c}{37.01\\0.934}
        &\tabincell{c}{26.19\\0.817}
        &\tabincell{c}{28.64\\0.853}
        &\tabincell{c}{26.47\\0.806}
        &\tabincell{c}{26.09\\0.831}
        &\tabincell{c}{27.50\\0.826}
        &\tabincell{c}{27.13\\0.816}
        &\tabincell{c}{28.53\\0.841}
        \\
        % \midrule
        % HSSP \cite{8954038}
        % & - 
        % & -
        % &\tabincell{c}{31.48\\0.858}
        % &\tabincell{c}{31.09\\0.842}
        % &\tabincell{c}{28.96\\0.823}
        % &\tabincell{c}{34.56\\0.902}
        % &\tabincell{c}{28.53\\0.808}
        % &\tabincell{c}{30.83\\0.877}
        % &\tabincell{c}{28.71\\0.824}
        % &\tabincell{c}{30.09\\0.881}
        % &\tabincell{c}{30.43\\0.868}
        % &\tabincell{c}{28.78\\0.842}
        % &\tabincell{c}{30.35\\0.852}
        % \\
        \midrule
        DNU \cite{9156942}
        & 1.19M
        & 163.48
        &\tabincell{c}{31.72\\0.863}
        &\tabincell{c}{31.13\\0.846}
        &\tabincell{c}{29.99\\0.845}
        &\tabincell{c}{35.34\\0.908}
        &\tabincell{c}{29.03\\0.833}
        &\tabincell{c}{30.87\\0.887}
        &\tabincell{c}{28.99\\0.839}
        &\tabincell{c}{30.13\\0.885}
        &\tabincell{c}{31.03\\0.876}
        &\tabincell{c}{29.14\\0.849}
        &\tabincell{c}{30.74\\0.863}
        \\
        \midrule
        DIP-HSI \cite{9710184}
        & 33.85M
        & 64.42
        &\tabincell{c}{32.68\\0.890}
        &\tabincell{c}{27.26\\0.833}
        &\tabincell{c}{31.30\\0.914}
        &\tabincell{c}{40.54\\0.962}
        &\tabincell{c}{29.79\\0.900}
        &\tabincell{c}{30.39\\0.877}
        &\tabincell{c}{28.18\\0.913}
        &\tabincell{c}{29.44\\0.874}
        &\tabincell{c}{34.51\\0.927}
        &\tabincell{c}{28.51\\0.851}
        &\tabincell{c}{31.26\\0.894}
        \\
        % \midrule
        % TSA-Net \cite{meng2020end}
        % & 44.25M
        % & 110.06
        % &\tabincell{c}{32.03\\0.892}
        % &\tabincell{c}{31.00\\0.858}
        % &\tabincell{c}{32.25\\0.915}
        % &\tabincell{c}{39.19\\0.953}
        % &\tabincell{c}{29.39\\0.884}
        % &\tabincell{c}{31.44\\0.908}
        % &\tabincell{c}{30.32\\0.878}
        % &\tabincell{c}{29.35\\0.888}
        % &\tabincell{c}{30.01\\0.890}
        % &\tabincell{c}{29.59\\0.874}
        % &\tabincell{c}{31.46\\0.894}
        % \\
        \midrule
        DGSMP \cite{huang2023deep}
        & 3.76M
        & 646.65
        &\tabincell{c}{33.26\\0.915}
        &\tabincell{c}{32.09\\0.898}
        &\tabincell{c}{33.06\\0.925}
        &\tabincell{c}{40.54\\0.964}
        &\tabincell{c}{28.86\\0.882}
        &\tabincell{c}{33.08\\0.937}
        &\tabincell{c}{30.74\\0.886}
        &\tabincell{c}{31.55\\0.923}
        &\tabincell{c}{31.66\\0.911}
        &\tabincell{c}{31.44\\0.925}
        &\tabincell{c}{32.63\\0.917}
        \\
        \midrule
        GAP-Net \cite{meng2020gap}
        & 4.27M
        & 78.58
        &\tabincell{c}{33.74\\0.911}
        &\tabincell{c}{33.26\\0.900}
        &\tabincell{c}{34.28\\0.929}
        &\tabincell{c}{41.03\\0.967}
        &\tabincell{c}{31.44\\0.919}
        &\tabincell{c}{32.40\\0.925}
        &\tabincell{c}{32.27\\0.902}
        &\tabincell{c}{30.46\\0.905}
        &\tabincell{c}{33.51\\0.915}
        &\tabincell{c}{30.24\\0.895}
        &\tabincell{c}{33.26\\0.917}
        \\
        \midrule
        ADMM-Net \cite{9009497}
        & 4.27M
        & 78.58
        &\tabincell{c}{34.12\\0.918}
        &\tabincell{c}{33.62\\0.902}
        &\tabincell{c}{35.04\\0.931}
        &\tabincell{c}{41.15\\0.966}
        &\tabincell{c}{31.82\\0.922}
        &\tabincell{c}{32.54\\0.924}
        &\tabincell{c}{32.42\\0.896}
        &\tabincell{c}{30.74\\0.907}
        &\tabincell{c}{33.75\\0.915}
        &\tabincell{c}{30.68\\0.895}
        &\tabincell{c}{33.58\\0.918}
        \\
        % \iffalse
        % \midrule
        % MST-S \cite{9857277}
        % & 0.93M
        % & 12.96
        % &\tabincell{c}{34.71\\0.930}
        % &\tabincell{c}{34.45\\0.925}
        % &\tabincell{c}{35.32\\0.943}
        % &\tabincell{c}{41.50\\0.967}
        % &\tabincell{c}{31.90\\0.933}
        % &\tabincell{c}{33.85\\0.943}
        % &\tabincell{c}{32.69\\0.911}
        % &\tabincell{c}{31.69\\0.933}
        % &\tabincell{c}{34.67\\0.939}
        % &\tabincell{c}{31.82\\0.926}
        % &\tabincell{c}{34.26\\0.935}
        % \\
        % \midrule
        % MST-M \cite{mst}
        % & 1.50M
        % & 18.07
        % &\tabincell{c}{35.15\\0.937}
        % &\tabincell{c}{35.19\\0.935}
        % &\tabincell{c}{36.26\\0.950}
        % &\tabincell{c}{{42.48}\\0.973}
        % &\tabincell{c}{32.49\\0.943}
        % &\tabincell{c}{34.28\\0.948}
        % &\tabincell{c}{33.29\\0.921}
        % &\tabincell{c}{32.40\\0.943}
        % &\tabincell{c}{35.35\\0.942}
        % &\tabincell{c}{32.53\\0.935}
        % &\tabincell{c}{34.94\\0.943}
        % \\
        % \fi
        \midrule
        HDNet \cite{9879297}
        & 2.37M
        & 154.76
        &\tabincell{c}{35.14\\0.935}
        &\tabincell{c}{35.67\\0.940}
        &\tabincell{c}{36.03\\0.943}
        &\tabincell{c}{42.30\\0.969}
        &\tabincell{c}{32.69\\0.946}
        &\tabincell{c}{34.46\\0.952}
        &\tabincell{c}{33.67\\0.926}
        &\tabincell{c}{32.48\\0.941}
        &\tabincell{c}{34.89\\0.942}
        &\tabincell{c}{32.38\\0.937}
        &\tabincell{c}{34.97\\0.943}
        \\
        \midrule
        MST-L \cite{cai2022mask}
        & 2.03M
        & 28.15
        &\tabincell{c}{35.40\\0.941}
        &\tabincell{c}{35.87\\0.944}
        &\tabincell{c}{36.51\\0.953}
        &\tabincell{c}{42.27\\0.973}
        &\tabincell{c}{32.77\\0.947}
        &\tabincell{c}{34.80\\0.955}
        &\tabincell{c}{33.66\\0.925}
        &\tabincell{c}{32.67\\0.948}
        &\tabincell{c}{35.39\\0.949}
        &\tabincell{c}{32.50\\0.941}
        &\tabincell{c}{35.18\\0.948}
        \\
        \midrule
        MST++ \cite{9857277}
        & 1.33M
        & 19.42
        &\tabincell{c}{35.80\\0.943}
        &\tabincell{c}{36.23\\0.947}
        &\tabincell{c}{37.34\\0.957}
        &\tabincell{c}{42.63\\0.973}
        &\tabincell{c}{33.38\\0.952}
        &\tabincell{c}{35.38\\0.957}
        &\tabincell{c}{34.35\\0.934}
        &\tabincell{c}{33.71\\0.953}
        &\tabincell{c}{36.67\\0.953}
        &\tabincell{c}{33.38\\0.945}
        &\tabincell{c}{35.99\\0.951}
        \\
        \midrule
        CST-L \cite{cai2022coarse}
        & 3.00M
        & 40.01
        &\tabincell{c}{35.96\\0.949}
        &\tabincell{c}{36.84\\0.955}
        &\tabincell{c}{38.16\\0.962}
        &\tabincell{c}{42.44\\0.975}
        &\tabincell{c}{33.25\\0.955}
        &\tabincell{c}{35.72\\0.963}
        &\tabincell{c}{34.86\\0.944}
        &\tabincell{c}{34.34\\0.961}
        &\tabincell{c}{36.51\\0.957}
        &\tabincell{c}{33.09\\0.945}
        &\tabincell{c}{36.12\\0.957}
        \\
        \midrule
        %\rowcolor{lightgray}
        BIRNAT~\cite{9741335}
        & 4.40M
        & 2122.66
        &\tabincell{c}{36.79\\0.951}
        &\tabincell{c}{37.89\\0.957}
        &\tabincell{c}{40.61\\0.971}
        &\tabincell{c}{46.94\\0.985}
        &\tabincell{c}{35.42\\0.964}
        &\tabincell{c}{35.30\\0.959}
        &\tabincell{c}{36.58\\0.955}
        &\tabincell{c}{33.96\\0.956}
        &\tabincell{c}{39.47\\0.970}
        &\tabincell{c}{32.80\\0.938}
        &\tabincell{c}{37.58\\0.960}
        \\
        \midrule
        LDMUN \cite{wu2023latent}
        & --
        & --
        &\tabincell{c}{38.07\\0.969}
        &\tabincell{c}{41.16\\0.982}
        &\tabincell{c}{43.70\\0.983}
        &\tabincell{c}{48.01\\0.993}
        &\tabincell{c}{37.76\\0.980}
        &\tabincell{c}{37.65\\0.980}
        &\tabincell{c}{38.58\\0.973}
        &\tabincell{c}{36.31\\0.979}
        &\tabincell{c}{42.66\\0.984}
        &\tabincell{c}{35.18\\0.967}
        &\tabincell{c}{39.91\\0.979}
        \\
        \midrule
        DAUHST \cite{cai2022degradationaware}
        & 6.15M
        & 79.50
        &\tabincell{c}{37.25\\0.958}
        &\tabincell{c}{39.02\\0.967}
        &\tabincell{c}{41.05\\0.971}
        &\tabincell{c}{46.15\\0.983}
        &\tabincell{c}{35.80\\0.969}
        &\tabincell{c}{37.08\\0.970}
        &\tabincell{c}{37.57\\0.963}
        &\tabincell{c}{35.10\\0.966}
        &\tabincell{c}{40.02\\0.970}
        &\tabincell{c}{34.59\\0.956}
        &\tabincell{c}{38.36\\0.967}
        \\
        \midrule
        PADUT \cite{li2023pixel}
        & 5.38M
        & 90.46
        &\tabincell{c}{37.36\\0.962}
        &\tabincell{c}{40.43\\0.978}
        &\tabincell{c}{42.38\\0.979}
        &\tabincell{c}{46.62\\0.990}
        &\tabincell{c}{36.26\\0.974}
        &\tabincell{c}{37.27\\0.974}
        &\tabincell{c}{37.83\\0.966}
        &\tabincell{c}{35.33\\0.974}
        &\tabincell{c}{40.86\\0.978}
        &\tabincell{c}{34.55\\0.963}
        &\tabincell{c}{38.89\\0.974}
        \\
        \midrule
        RDLUF \cite{dong2023residual}
        & 1.89M
        & 115.34
        &\tabincell{c}{37.94\\0.966}
    &\tabincell{c}{40.95\\0.977}
    &\tabincell{c}{43.25\\0.979}
    &\tabincell{c}{47.83\\0.990}
    &\tabincell{c}{37.11\\0.976}
    &\tabincell{c}{37.47\\0.975}
    &\tabincell{c}{38.58\\0.969}
    &\tabincell{c}{35.50\\0.970}
    &\tabincell{c}{41.83\\0.978}
    &\tabincell{c}{35.23\\0.962}
    &\tabincell{c}{39.57\\0.974}
        \\
        \midrule
        DERNN (3stg) \cite{dernn_lnlt}
            & 0.65M
            & 27.41
            &\tabincell{c}{37.54 \\ 0.964}
            &\tabincell{c}{39.23 \\ 0.973}
            &\tabincell{c}{42.01 \\ 0.979}
            &\tabincell{c}{47.08 \\ 0.992}
            &\tabincell{c}{36.03 \\ 0.973}
            &\tabincell{c}{36.82 \\ 0.974}
            &\tabincell{c}{37.34 \\ 0.966}
            &\tabincell{c}{35.04 \\ 0.971}
            &\tabincell{c}{40.97 \\ 0.978}
            &\tabincell{c}{34.39 \\ 0.960}
            &\tabincell{c}{38.65 \\ 0.973}
            \\
        \midrule    
        DERNN (5stg) \cite{dernn_lnlt}
            & 0.65M
            & 45.60
            &\tabincell{c}{37.86 \\ 0.963}
            &\tabincell{c}{40.28 \\ 0.976}
            &\tabincell{c}{42.69 \\ 0.978}
            &\tabincell{c}{47.97 \\ 0.990}
            &\tabincell{c}{37.11 \\ 0.975}
            &\tabincell{c}{37.23 \\ 0.974}
            &\tabincell{c}{37.97 \\ 0.967}
            &\tabincell{c}{35.82 \\ 0.971}
            &\tabincell{c}{41.93 \\ 0.979}
            &\tabincell{c}{34.98 \\ 0.959}
            &\tabincell{c}{39.38 \\ 0.973}
            \\
            \midrule
            DERNN (7stg) \cite{dernn_lnlt}
            & 0.65M
            & 63.80
            % &\tabincell{c}{PSNR $\uparrow$ \\ SSIM $\uparrow$ \\ SAM $\downarrow$}
            &\tabincell{c}{37.91 \\ 0.964}
            &\tabincell{c}{40.75 \\ 0.978}
            &\tabincell{c}{42.95 \\ 0.978}
            &\tabincell{c}{47.51 \\ 0.990}
            &\tabincell{c}{37.81 \\ 0.978} 
            &\tabincell{c}{37.37 \\ 0.975}
            &\tabincell{c}{38.49 \\ 0.970}
            &\tabincell{c}{35.83 \\ 0.971}
            &\tabincell{c}{42.47 \\ 0.980}
            &\tabincell{c}{35.04 \\ 0.961} 
            &\tabincell{c}{39.61 \\ 0.974}
            \\
            \midrule
            DERNN (9stg) \cite{dernn_lnlt}
            & 0.65M
            & 81.99
            % &\tabincell{c}{PSNR $\uparrow$ \\ SSIM $\uparrow$ \\ SAM $\downarrow$}
            &\tabincell{c}{38.26 \\ 0.965}
            &\tabincell{c}{40.97 \\ 0.979}
            &\tabincell{c}{43.22 \\ 0.979}
            &\tabincell{c}{48.10 \\ 0.991}
            &\tabincell{c}{38.08 \\ 0.980}
            &\tabincell{c}{37.41 \\ 0.975}
            &\tabincell{c}{38.83 \\ 0.971}
            &\tabincell{c}{36.41 \\ 0.973}
            &\tabincell{c}{42.87 \\ 0.981}
            &\tabincell{c}{35.15 \\ 0.962}
            &\tabincell{c}{39.93 \\ 0.976}
            \\
            \midrule
            % \rowcolor{rouse}
            DERNN (9stg$^*$) \cite{dernn_lnlt}
            & 1.09M
            & 134.18
            &\tabincell{c}{38.49 \\ 0.968} 
            &\tabincell{c}{41.27 \\ 0.980} 
            &\tabincell{c}{ 43.97 \\ 0.980} 
            &\tabincell{c}{\bf 48.61 \\ 0.992} 
            &\tabincell{c}{38.29 \\ 0.981} 
            &\tabincell{c}{37.81 \\ 0.977} 
            &\tabincell{c}{39.30 \\ 0.973} 
            &\tabincell{c}{36.51 \\ 0.974} 
            &\tabincell{c}{43.38 \\ 0.983} 
            &\tabincell{c}{35.61 \\ 0.966}
            &\tabincell{c}{40.33 \\ 0.977} 
            \\
        \midrule
        \rowcolor{rouse}
        \bf DHM-light (3stg) 
        & 0.66M
        & 26.42
        & \tabincell{c}{37.67 \\ 0.965}
        &\tabincell{c}{39.58 \\ 0.974} 
        &\tabincell{c}{42.67 \\ 0.981} 
        &\tabincell{c}{47.90 \\ 0.993} 
        &\tabincell{c}{36.47 \\ 0.975} 
        &\tabincell{c}{36.76 \\ 0.975} 
        &\tabincell{c}{37.72 \\ 0.968} 
        &\tabincell{c}{35.14 \\ 0.972} 
        &\tabincell{c}{41.65 \\ 0.981} 
        &\tabincell{c}{34.35 \\ 0.961}
        &\tabincell{c}{38.99 \\ 0.975}
        \\
        \midrule
        \rowcolor{rouse}
        \bf DHM-light (5stg) 
        & 0.66M
        & 43.96
         & \tabincell{c}{38.17 \\ 0.971}
        &\tabincell{c}{40.91 \\ 0.981} 
        &\tabincell{c}{43.78 \\ 0.983} 
        &\tabincell{c}{47.18 \\ 0.993} 
        &\tabincell{c}{37.41 \\ 0.980} 
        &\tabincell{c}{37.51 \\ 0.978} 
        &\tabincell{c}{38.78 \\ 0.973} 
        &\tabincell{c}{35.83 \\ 0.977} 
        &\tabincell{c}{43.26 \\ 0.985} 
        &\tabincell{c}{35.28 \\ 0.968}
        &\tabincell{c}{39.81 \\ 0.979}
        \\
        \midrule
        \rowcolor{rouse}
        \bf DHM-light (7stg) 
        &  0.66M
        & 61.50
        & \tabincell{c}{38.58 \\ 0.972}
        &\tabincell{c}{41.42 \\ 0.983} 
        &\tabincell{c}{43.93 \\ 0.984} 
        &\tabincell{c}{47.95 \\ 0.993} 
        &\tabincell{c}{38.29 \\ 0.983} 
        &\tabincell{c}{37.88 \\ 0.980} 
        &\tabincell{c}{39.03 \\ 0.974} 
        &\tabincell{c}{36.26 \\ 0.979} 
        &\tabincell{c}{43.25 \\ 0.986} 
        &\tabincell{c}{35.42 \\ 0.970}
        &\tabincell{c}{40.20 \\ 0.980}
        \\
        \midrule
        \rowcolor{rouse}
        \bf DHM-light (9stg) 
        & 0.66M
        & 79.04
        & \tabincell{c}{ \bf 38.78 \\ \bf0.972}
        &\tabincell{c}{41.44 \\ 0.983} 
        &\tabincell{c}{44.07 \\ 0.984} 
        &\tabincell{c}{48.16 \\ \bf 0.994} 
        &\tabincell{c}{38.32 \\ 0.983} 
        &\tabincell{c}{37.45 \\ 0.980} 
        &\tabincell{c}{39.22 \\ 0.976} 
        &\tabincell{c}{36.37 \\ 0.980} 
        &\tabincell{c}{43.75 \\ 0.987} 
        &\tabincell{c}{35.73 \\ 0.972}
        &\tabincell{c}{40.33 \\ 0.981}
        \\
        \midrule
        \rowcolor{rouse}
        \bf DHM (3stg) 
        & 0.92M
        & 36.34
        & \tabincell{c}{37.63 \\ 0.967}
        &\tabincell{c}{39.85 \\ 0.976} 
        &\tabincell{c}{43.40 \\ 0.982} 
        &\tabincell{c}{47.56 \\ 0.993} 
        &\tabincell{c}{36.37 \\ 0.976} 
        &\tabincell{c}{36.98 \\ 0.975} 
        &\tabincell{c}{38.05 \\ 0.970} 
        &\tabincell{c}{34.94 \\ 0.972} 
        &\tabincell{c}{42.04 \\ 0.982} 
        &\tabincell{c}{34.42 \\ 0.962}
        &\tabincell{c}{39.13 \\ 0.975}
        \\
        \midrule
        \rowcolor{rouse}
        \bf DHM (5stg) 
        & 0.92M
        & 60.50
        & \tabincell{c}{38.48 \\ 0.972}
        &\tabincell{c}{41.14 \\ 0.982} 
        &\tabincell{c}{44.10 \\ 0.984} 
        &\tabincell{c}{48.03 \\ 0.993} 
        &\tabincell{c}{37.82 \\ 0.981} 
        &\tabincell{c}{37.95 \\ 0.979} 
        &\tabincell{c}{39.21 \\ 0.975} 
        &\tabincell{c}{36.34 \\ 0.978} 
        &\tabincell{c}{43.31 \\ 0.986} 
        &\tabincell{c}{35.20 \\ 0.967}
        &\tabincell{c}{40.16 \\ 0.980}
        \\
        \midrule
        \rowcolor{rouse}
        \bf DHM (7stg) 
        & 0.92M
        & 84.65
        & \tabincell{c}{38.40 \\ 0.972}
        &\tabincell{c}{41.52 \\ 0.983} 
        &\tabincell{c}{44.21 \\ 0.984} 
        &\tabincell{c}{47.93 \\ 0.994} 
        &\tabincell{c}{38.21 \\ 0.983} 
        &\tabincell{c}{38.17 \\ 0.981} 
        &\tabincell{c}{39.58 \\ 0.976} 
        &\tabincell{c}{36.17 \\ 0.978} 
        &\tabincell{c}{43.56 \\ 0.986} 
        &\tabincell{c}{35.60 \\ 0.970}
        &\tabincell{c}{40.34 \\ 0.981}
        \\
        \midrule
        \rowcolor{rouse}
        \bf DHM (9stg) 
        & 0.92M
        & 108.80
        & \tabincell{c}{38.50 \\  0.972}
        &\tabincell{c}{\bf 41.64 \\ \bf 0.984} 
        &\tabincell{c}{\bf 44.37 \\ \bf 0.985} 
        &\tabincell{c}{48.13 \\  0.994} 
        &\tabincell{c}{\bf 38.33 \\ \bf 0.983} 
        &\tabincell{c}{\bf 38.27 \\ \bf 0.982} 
        &\tabincell{c}{\bf 39.70 \\ \bf 0.977} 
        &\tabincell{c}{\bf 36.52 \\ \bf 0.980} 
        &\tabincell{c}{\bf 43.89 \\ \bf 0.988} 
        &\tabincell{c}{\bf 35.75 \\ \bf 0.971}
        &\tabincell{c}{\bf 40.50 \\ \bf 0.982}
        \\
        \bottomrule
    \end{tabular}
}
\label{tab: comparison_simu}
\vspace{-15pt}
\end{table*}

\vspace{-2mm}
\subsection{Quantitative Performance Comparisons}
\vspace{-2mm}
As shown in Tab.~\ref{tab: comparison_simu}, we introduce comprehensive quantitative comparisons between our HDM and SOTA HSI reconstruction methods on the simulation dataset with 10 scenes (S1$\sim$S10). From the results in Tab.~\ref{tab: comparison_simu}, we observe that the proposed DHM (9stg) (\emph{i.e.}, our DHM at the $9$-th stage) achieves the best HSI reconstruction performance (\emph{i.e.}, 40.50 dB in PSNR and 0.982 in SSIM). Our DHM (9stg) substantially surpasses existing methods \cite{4358846, 8481592, 9710184, cai2022coarse, huang2021deep}, especially several recent SOTA comparison models (\emph{e.g.}, DAUHST \cite{cai2022degradationaware}, LDMUN \cite{wu2023latent}, RDLUF-Mix \cite{dong2023residual}, DERNN  \cite{dernn_lnlt}) by $0.57\sim2.14$ dB. Such improvements verify the effectiveness of our DHM in exploring long-range dependencies across the entire high-resolution HSIs using global receptive fields, while capturing local context within local windows. More importantly, our DHM requires lower model size and computational costs to dramatically outperform existing methods. Compared with the SOTA DERNN (9stg$^{*}$) \cite{dernn_lnlt}, our DHM (9stg) improves 0.17 dB in PSNR and 0.005 in SSIM, but only consumes 84.40\% (0.92M / 1.09M) parameters and 81.09\% (108.80 / 134.18) GFLOPS. Moreover, we propose a light model (\emph{i.e.}, DHM-light) where each DHSB contains a single global hyperspectral S4 block (GHSB) and a GFFN. In Tab.~\ref{tab: comparison_simu}, our DHM-light at the 3/5/7/9-th stage has significant improvement than other comparison methods (\emph{e.g.}, DERNN \cite{dernn_lnlt}) with the same number of stages, while retaining comparable model size and less GFLOPs. It illustrate the effectiveness of our DHM for HSI reconstruction task.

\begin{figure*}[t]
\centering
\includegraphics[width=.99999\linewidth]
{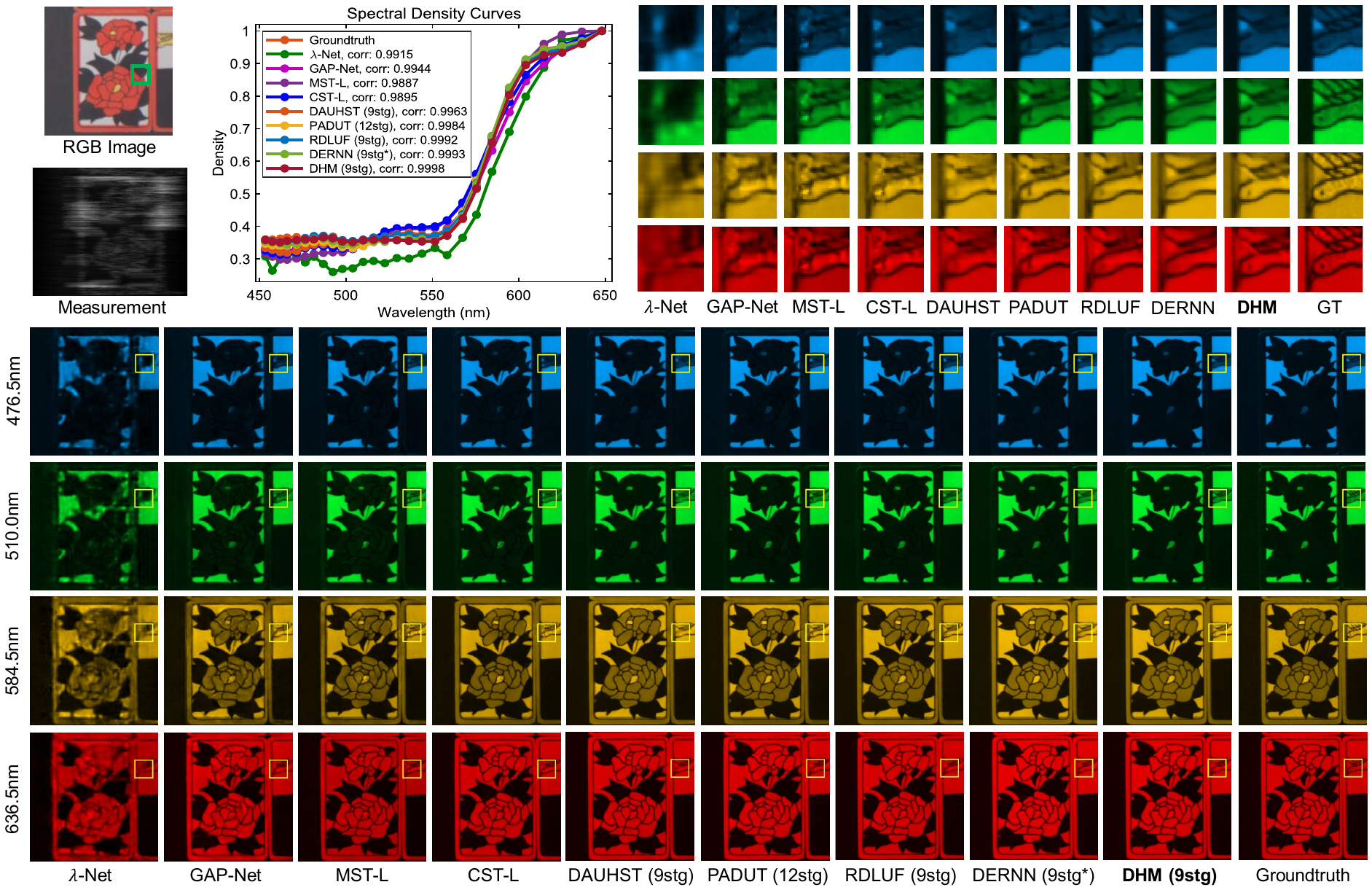}
\vspace{-6mm}
\caption{Qualitative results on the Scene 7 (S7) of simulation dataset (zoom in for a better view).}
\label{fig: vis_simu}
\vspace{-5mm}
\end{figure*}

\begin{figure*}[t]
\centering
\includegraphics[width=.99999\linewidth]
{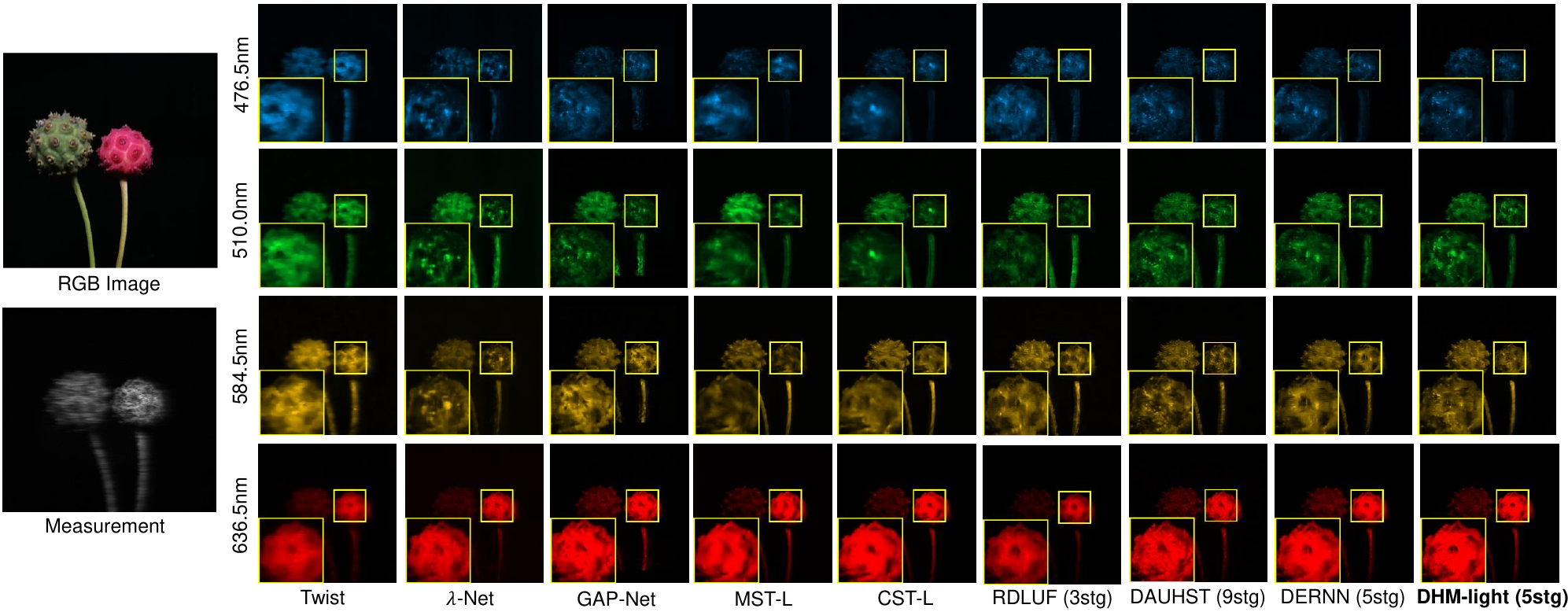}
\vspace{-6mm}
\caption{Qualitative comparisons on the Scene 4 (S4) of real dataset (zoom in for a better view).}
\label{fig: vis_real}
\vspace{-3mm}
\end{figure*}

\vspace{-2mm}
\subsection{Qualitative Performance Comparisons}
\vspace{-2mm}
\textbf{Simulation Dataset:} 
As depicted in Fig.~\ref{fig: vis_simu}, we select 4 out of the 28 spectral channels to visualize some qualitative comparisons of HSI reconstruction on the Scene 7 (S7) of simulation dataset. For better visibility, we zoom in on the regions within the yellow boxes of the original HSIs (bottom), and show the comparison of these regions in the top-right part. In Fig.~\ref{fig: vis_simu}, previous methods suffer from blotchy texture, distortions and blurring artifacts. In contrast, our DHM (9stg) can effectively restore HSIs with less artifacts and finer details. 
Besides, the spectral density curves corresponding to the green  boxes in the top-left RGB image are depicted in the top-middle part. Our DHM (9stg) exhibits the best correlation with groundtruth, which illustrates the effectiveness of our DHM.

\textbf{Real Dataset:}
To verify the superiority of our model in real HSI reconstruction, we follow \cite{meng2020end, cai2022degradationaware, wu2023latent, dernn_lnlt} to retrain our DHM-light (5stg) on the joint KAIST \cite{kaist} and CAVE \cite{cave}. Besides, we introduce 11-bit shot noise into training samples to simulate real imaging scenarios. As shown in Fig.~\ref{fig: vis_real}, our DHM-light (5stg) can effectively restore the plant region corresponding to the yellow box. Compared with SOTA methods \cite{cai2022degradationaware, li2023pixel, dernn_lnlt}, our DHM-light (5stg) restores clearer contents and structural details with less artifacts, verifying the robustness of our model to address the real HSI restoration.

\begin{table*}[t]
\caption{Ablation studies (averaged PSNR and SSIM) of our DHM (5stg) on simulation dataset.}
\vspace{-2mm}
\centering
\subfloat[\footnotesize Ablation experiments of the DHSB.\label{abl:1}]
{
        \scalebox{0.77}{
        \setlength{\tabcolsep}{1.1mm}
        \begin{tabular}{c c c| c c c c c}
        \toprule
        GHSB & LHSB & GFFN & \#Params & GFLOPs & PSNR & SSIM \\
        \midrule
        \checkmark & & \checkmark& 0.66M & 43.96 & 39.81 & 0.979\\
        &\checkmark & \checkmark & 0.66M & 43.96 & 38.76 & 0.973\\
        \checkmark & \checkmark &  & 0.92M & 60.26  & 39.93 & 0.979 \\
        \checkmark & \checkmark& \checkmark & 0.92M & 60.50 & \bf 40.16 & \bf 0.980 \\
        \bottomrule
\end{tabular}}} \hspace{0mm}
\subfloat[\footnotesize  Ablation experiments of alternative variants.\label{abl:2}]{
        \scalebox{0.77}{
        \setlength{\tabcolsep}{1.1mm}
        \begin{tabular}{c c|c c c c c}
        \toprule
         GHSB$\rightarrow$GA & LHSB$\rightarrow$LA & \#Params & GFLOPs & PSNR & SSIM \\
        \midrule
        & & 0.92M & 60.50 & \bf 40.16 & \bf 0.980 \\
        \checkmark &  & 0.79M  & 53.05 & 39.11  & 0.975 \\
         & \checkmark & 0.79M  & 53.05 & 40.08  & 0.980 \\
         \checkmark & \checkmark & 0.65M & 45.60 & 39.38 &  0.973 \\
        \bottomrule
\end{tabular}}}\vspace{-3mm}\\

\subfloat[\footnotesize Ablation analysis of different block orders.\label{abl:3}]
{
        \scalebox{0.77}{
        \setlength{\tabcolsep}{1.2mm}
        \begin{tabular}{c c c| c c c c c}
        \toprule
        GS$\rightarrow$LS & LS$\rightarrow$GS & SPs & \#Params & GFLOPs & PSNR & SSIM \\
        \midrule
         & \checkmark & & 4.59M & 60.50 & 39.23 & 0.977 \\
         & \checkmark & \checkmark & 0.92M & 60.50 & 40.12 & 0.980 \\
         \checkmark & &  & 4.59M & 60.50 & 39.28 & 0.977 \\
        \checkmark & & \checkmark & 0.92M & 60.50 & \bf 40.16 & \bf 0.980 \\
        \bottomrule
\end{tabular}}} \hspace{0mm}
\subfloat[\footnotesize Ablation analysis of learnable $(\boldsymbol{\eta}, \boldsymbol{\rho})$.\label{abl:4}]{
\scalebox{0.77}{
        \setlength{\tabcolsep}{1.1mm}
        \begin{tabular}{l|c c|c c c c c}
        \toprule
         \makecell[c]{Variants} & $ ~~\boldsymbol{\eta}$~~ & $~~\boldsymbol{\rho}$~~ & \#Params & GFLOPs & PSNR & SSIM \\
        \midrule
         Baseline &  &  & 0.90M &  59.11 & 39.86 & 0.979\\
         DHM w/o $\boldsymbol{\eta}$ &  & \checkmark & 0.92M & 60.50  & 39.71 & 0.978\\
         DHM w/o $\boldsymbol{\rho}$ & \checkmark &  &  0.92M & 60.42 &  39.92 & 0.979 \\
         DHM & \checkmark &\checkmark & 0.92M & 60.50 & \bf 40.16 & \bf 0.980 \\
        \bottomrule
\end{tabular}}}\hspace{0mm}
\label{table:ablation}
\vspace{-7mm}
\end{table*}

\begin{table*}[t]
\caption{Ablation results of the DHSB. In each cell, the upper and lower entries are PSNR and SSIM.}
\vspace{-2mm}
\centering
\resizebox{0.99999\textwidth}{!}
{
    % \centering
    % \begin{tabular}{c|c|c|c|c|c|c|>{\columncolor{lightgray}}c}
    \begin{tabular}{ccc|cc|cccccccccc|c}
        \toprule
        % \rowcolor{lightgray}
        GHSB & LHSB & GFFN
        &\#Params
        &GFLOPS
        & ~S1~
        & ~S2~
        & ~S3~
        & ~S4~
        & ~S5~
        & ~S6~
        & ~S7~
        & ~S8~
        & ~S9~
        & ~S10~
        & ~Avg~
        \\
        \midrule
        \checkmark & & \checkmark
        & 0.66M 
        & 43.96
        & \tabincell{c}{38.17 \\ 0.971}
        &\tabincell{c}{40.91 \\ 0.981} 
        &\tabincell{c}{43.78 \\ 0.983} 
        &\tabincell{c}{47.18 \\ 0.993} 
        &\tabincell{c}{37.41 \\ 0.980} 
        &\tabincell{c}{37.51 \\ 0.978} 
        &\tabincell{c}{38.78 \\ 0.973} 
        &\tabincell{c}{35.83 \\ 0.977} 
        &\tabincell{c}{43.26 \\ 0.985} 
        &\tabincell{c}{\bf35.28 \\ \bf0.968}
        &\tabincell{c}{39.81 \\ 0.979}
        \\
        \midrule
        & \checkmark & \checkmark
        & 0.66M 
        & 43.96
        & \tabincell{c}{37.28 \\ 0.962}
        &\tabincell{c}{39.95 \\ 0.975} 
        &\tabincell{c}{42.77 \\ 0.981} 
        &\tabincell{c}{47.42 \\ 0.992} 
        &\tabincell{c}{35.95 \\ 0.973} 
        &\tabincell{c}{36.65 \\ 0.974} 
        &\tabincell{c}{37.40 \\ 0.966} 
        &\tabincell{c}{34.94 \\ 0.971} 
        &\tabincell{c}{41.00 \\ 0.979} 
        &\tabincell{c}{34.28 \\ 0.960}
        &\tabincell{c}{38.76 \\ 0.973}
        
        \\
        \midrule
        \checkmark & \checkmark & 
        & 0.92M 
        & 60.26
	& \tabincell{c}{38.42 \\ 0.972}
	&\tabincell{c}{40.75 \\ 0.981} 
	&\tabincell{c}{43.97 \\ 0.984} 
	&\tabincell{c}{47.65 \\ 0.993} 
	&\tabincell{c}{37.79 \\ 0.981} 
	&\tabincell{c}{37.47 \\ 0.978} 
	&\tabincell{c}{38.95 \\ 0.974} 
	&\tabincell{c}{35.96 \\ 0.976} 
	&\tabincell{c}{43.18 \\ 0.985} 
	&\tabincell{c}{35.13 \\ 0.967}
	&\tabincell{c}{39.93 \\ 0.979}
    
        \\
        \midrule
        \checkmark & \checkmark & \checkmark
        & 0.92M
        & 60.50
        & \tabincell{c}{\bf 38.48 \\ \bf 0.972}
        &\tabincell{c}{\bf 41.14 \\ \bf 0.982} 
        &\tabincell{c}{\bf 44.10 \\ \bf 0.984} 
        &\tabincell{c}{\bf 48.03 \\ \bf 0.993} 
        &\tabincell{c}{\bf 37.82 \\ \bf 0.981} 
        &\tabincell{c}{\bf 37.95 \\ \bf 0.979} 
        &\tabincell{c}{\bf 39.21 \\ \bf 0.975} 
        &\tabincell{c}{\bf 36.34 \\ \bf 0.978} 
        &\tabincell{c}{\bf 43.31 \\ \bf 0.986} 
        &\tabincell{c}{ 35.20 \\  0.967}
        &\tabincell{c}{\bf 40.16 \\ \bf 0.980}
        \\
        \bottomrule
    \end{tabular}
}
\label{tab: ablation_model_more}
\vspace{-10pt}
\end{table*}

\vspace{-2mm}
\subsection{Ablation Studies}
\vspace{-2mm}
\begin{wrapfigure}{r}{0.43\textwidth}
\vspace{-11mm}
\begin{center} \hspace{-1.5mm}
\includegraphics[width=0.43\textwidth]{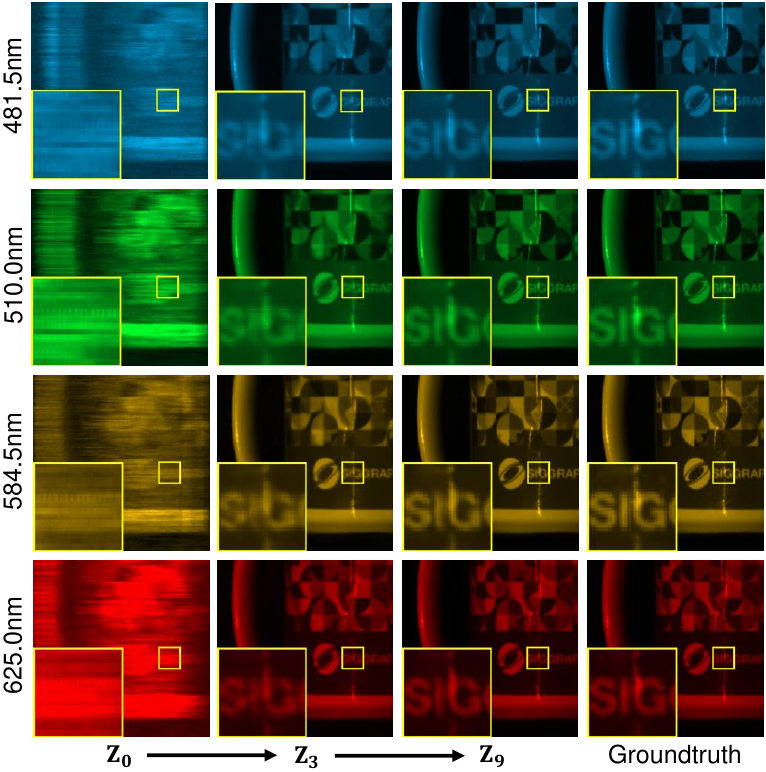}
\end{center}
\vspace{-4mm}
\caption{\small Visualization of $\mathbf{Z}_t$ at different stages on the Scene 5 (S5) of simulation dataset. }
\vspace{-4mm}
\label{fig: vis_z_t}
\end{wrapfigure}

This subsection analyzes the effectiveness of all proposed modules on simulation dataset using our DHM (5stg) as an example. 
\textbf{1) DHSB:} As shown in Tab.~\ref{abl:1}, when we remove the GHSB, LHSB or replace the GFFN with a traditional feed-forward network (FNN) \cite{cai2022degradationaware} in the DHSB, the performance of our DHM (5stg) significantly decreases by $0.23\sim1.40$ dB in PSNR and $0.001\sim0.007$ in SSIM. 
Tab.~\ref{tab: ablation_model_more} presents ablation results of our DHM (5stg) on 10 scenes (S1$\sim$S10) to veirify the effectiveness of the DHSB. 
\textbf{2) Variants:} In Tab.~\ref{abl:2}, our model decreases by $0.08\sim1.05$ dB in PSNR when we replace the GHSB with non-local MSA \cite{dernn_lnlt} (GHSB$\rightarrow$GA) or substitute the LHSB with local MSA \cite{dernn_lnlt} (LHSB$\rightarrow$LA), where MSA is the multi-head self-attention \cite{dosovitskiy2021an}. 
It verifies the effectiveness of our DHM in using global receptive fields to model long-range dependencies while capturing local contexts. 
\textbf{3) Block Orders:} In Tab.~\ref{abl:3}, we perform ablation studies about shared parameters (SPs) across different stages, and the orders of GHSB and LHSB: from GHSB to LHSB (GS$\rightarrow$LS) or from LHSB to GHSB (LS$\rightarrow$GS). The ablation results validate the effectiveness of our DHM. 
\textbf{4) Parameters:} Tab.~\ref{abl:4} shows ablation studies about learnable parameters $(\boldsymbol{\eta}, \boldsymbol{\rho})$, which validates their effectiveness to estimate degradation patterns. Fig.~\ref{fig: vis_z_t} visualizes $\{\mathbf{Z}_0, \mathbf{Z}_3, \mathbf{Z}_9\}$ as examples to verify the effectiveness of our unfolding framework in HSI reconstruction, when we use our DHM (9stg) as the denoiser $\mathcal{D}(\cdot)$.

\vspace{-3mm}
\section{Conclusion}
\vspace{-3mm}
In this paper, we propose a novel Dual Hyperspectral Mamba (DHM) to
model both global and local dependencies for efficient HSI reconstruction. After estimating degradation patterns of the CASSI system via the learnable parameters, we utilize these parameters to scale the linear projection and offer noise level for the denoiser (\emph{i.e.}, our DHM) in the multi-stage unfolding framework.  Particularly, the proposed DHM mainly consists of a global hyperspectral S4 block (GHSB) and a local hyperspectral S4 block (LHSB). The GHSB can explore long-range dependencies across the entire high-resolution HSIs using global receptive fields, while the LHSB constructs S4 models within different local windows to capture local contexts. We conduct enormous quantitative and qualitative comparison experiments on both the simulation and real datasets to demonstrate the effectiveness of our DHM.

%%%%%%%%%%%%%%%%%%%%%%%%%%%%%%%%%%%%%%%%%%%%%%%%%%%%%%%%%%%%
{\small
	\bibliographystyle{plain}
	\bibliography{DHM_arXiv}
}
%%%%%%%%%%%%%%%%%%%%%%%%%%%%%

\end{document}